\documentclass[letter]{aa} 

\usepackage[normalem]{ulem}

\usepackage{graphicx}
\usepackage{txfonts}
\usepackage{longtable}
\usepackage{amsmath}

\begin{document}

\title{Quantifying the evidence for resonant damping of coronal waves with foot-point wave power asymmetry}
\titlerunning{ResonantCoMP}

\author{M. Montes-Solís
	\inst{1,2,}\thanks{now at Space Weather Group, Departamento de Física y Matemáticas, Universidad de Alcalá, 28801 Madrid, Spain  (\email{m.montes@uah.es})}
	\and
	I. Arregui \inst{1,2}\
}

\institute{Instituto de Astrofísica de Canarias, E-38205 La Laguna, Tenerife, Spain
	\and
Departamento de Astrofísica, Universidad de La Laguna, E-38206 La Laguna, Tenerife, Spain	
}


\abstract {We use Coronal Multi-channel Polarimeter (CoMP) observations of propagating waves in the solar corona and Bayesian analysis to assess the evidence of models with resonant damping and foot-point wave power asymmetries. Two nested models are considered. The reduced model considers resonant damping as the sole cause of the measured discrepancy between outward and inward wave power. The larger model contemplates an extra source of asymmetry with origin at the foot-points. We first compute probability distributions of parameters conditional on the models and the observed data. The obtained constraints are then used to calculate the evidence for each model in view of data. We find that we need to consider the larger model to explain CoMP data and to accurately infer the damping ratio, hence, to better assess the possible contribution of the waves to coronal heating.}


\maketitle
  %
  
\section{Introduction}

The existence of propagating waves in extended regions of the solar corona was first demonstrated by \citet{Tomczyk2007} in observations taken with the Coronal Multi-channel Polarimeter (CoMP). These Doppler velocity fluctuations do not exhibit significant intensity variations and were initially interpreted as Alfv\'en waves propagating along the coronal magnetic field. Posterior theoretical results \citep{Goossens2012a} remark that a description in terms of resonantly damped kink waves offers a more accurate description for seismology and energy budget analyses. Resonant absorption is a plausible mechanism for explaining the damping of coronal MHD waves  \citep{Yo2017}.

The origin of the waves is uncertain.  Under some conditions, waves in the lower atmosphere can transport sufficient energy fluxes to the upper chromosphere or the overlying corona \citep[see e.g,][]{Jess2009,Srivastava2017,Liu2019}. The observed waves could  possibly contribute to the heating of the corona \citep{Arregui2015}. Although the measured Doppler velocity fluctuations have small amplitudes of the order of  $\sim$ 0.5 km s$^{-1}$ \citep{Tomczyk2007,Tomczyk2009,Morton2015}, the superposition of several oscillating structures along the line-of-sight can lead to an underestimation of the wave energy, with a large fraction being possibly hidden in the non-thermal line widths \citep{McIntosh2012,Pant2019}.

The observed waves are also relevant in solar atmospheric seismology. They show signatures of in situ wave damping in the form of a discrepancy between the outward and inward wave power \citep{Tomczyk2009,Morton2015}. Theoretical results for resonantly damped kink waves predict a frequency dependent damping. In the thin tube and thin boundary approximations, the damping length is inversely proportional to the frequency, thus high frequency waves are damped in shorter spatial scales \citep{Terradas2010}. 

\citet{Verth2010} produced a simple model to connect the damping rate of resonantly damped waves with the average power ratio for inward and outward propagating waves integrated along wave paths in an extended region of the corona. The model predicts an exponential dependence of the average power ratio with frequency. 
A least squares fit to the data shows a good qualitative agreement between the theoretical prediction and CoMP data. In a recent analysis, \citet{Tiwari2019} employ power ratio measurements along different paths which seem to confirm the frequency-dependent character of the damping.

Fitting a model to data consists of adopting a model $M$ and obtaining the set of so-called best fit parameters $\mbox{\boldmath$\theta$}$. In Bayesian inference the solution is given by the probability distribution of the parameters given the model and also conditional on data, $p(\mbox{\boldmath$\theta$}| M,D)$. Measuring the evidence for a model only makes sense in terms of its relative plausibility to another. An evidence analysis thus requires the assessment of each model plausibility conditional on data, $p(M|D)$, in addition to the calculation of the posterior probability distribution of the parameters.
	
The theoretical description by \cite{Verth2010} contains two possible causes for the observed average power ratio, resonant damping of propagating waves and the asymmetry in wave power at the foot-points.
	
In this Letter, we use CoMP data and two models to infer posterior probability distributions for the damping ratio due to resonant absorption and the power ratio at the foot-points.  Eventually, two sets of solutions equally well explain the data. A model comparison approach is used to quantify the evidence for each of them.

\section{Observed data and theoretical modelling}\label{model}

Sequences of Doppler images obtained with the CoMP instrument have permitted to perform time-distance seismology of the solar corona \citep{Tomczyk2007}. The procedure is based on constructing $k$-$\omega$ diagrams in extended regions of the corona from which outward and inward propagating waves can be identified and their relative amount of wave power measured. We consider CoMP observations taken on 2005 October 30. Fig.~4c in \citet{Tomczyk2009} displays the average $k$-$\omega$ diagram for a region around a sample wave path used to construct the wave diagnostics. The power ratio along the white dashed lines in Fig.~4c was further analysed by \citet{Verth2010} and is the basis of our study.  The data consist of a sample of average power ratio values at given frequencies in the range 0.05 to 4 mHz. 

\begin{figure}
	\centering
	\includegraphics[scale=0.7]{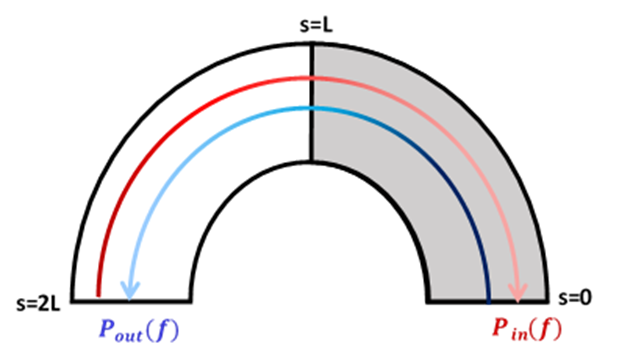}
	\caption{Model described by \citet{Verth2010}. Waves generated at the foot-points s=0 and s=2L propagate in opposite directions and are damped by resonant absorption in the whole annulus region. In the region of interest (grey), if $P_{\rm out}(f) = P_{\rm in}(f)$, waves in the inward direction (red) are more strongly damped than those in the outward direction (blue), as indicated by the coloured arrows.\label{f1}}
\end{figure}

The theoretical model by \citet{Verth2010} considers waves generated at two opposite foot-points propagating along a semi-circular region of total length $2L$ (Fig.~\ref{f1}). The waves are damped by resonant absorption and their amplitude decays exponentially as
\begin{equation}
A(s)=A_{0}\exp\left(-s/L_{\rm D}\right),
\end{equation}
with $L_{\rm D}$ the damping length. Under the thin tube and thin boundary approximations (TTTB), the damping length is inversely proportional to the wave frequency following the relation obtained by \cite{Terradas2010}
\begin{equation}
L_{D}=v_{\rm ph}\xi_{\rm E}\frac{1}{f},
\end{equation}
with $f=v_{\rm ph}/\lambda$ the frequency, $v_{\rm ph}$ the phase speed and $\lambda$ the wavelength. The parameter $\xi_{E}$ is the damping ratio, the ratio of damping length to the wavelength. For a sinusoidal variation of density in a non-uniform layer of length $l$ at the boundary of a waveguide with mean radius $R$, the damping ratio is \citep{Terradas2010}
\begin{equation}
\xi_E = \frac{L_{\rm D}}{\lambda}= \frac{2}{\pi} \left(\frac{\zeta + 1}{\zeta -1 }\right) \frac{R}{l},
\end{equation}
 with $\zeta=\rho_{\rm i}/\rho_{\rm e} > 1$ the density contrast. The damping ratio, $\xi_{\rm E}$,  is our first relevant parameter. Its possible values are in the interval $\xi_{\rm E}\in[1/\pi,\infty)$. The lower bound corresponds to the strong damping limit (for $l/R=2$, $\zeta\rightarrow\infty$). The upper bound to the case of no damping (for $l/R=0$).

\citet{Verth2010} obtained a simple analytical expression for the average outward to inward wave power ratio integrated along the wave path from $s=0$ to $s=L$ (shaded region in Fig.~\ref{f1}). This expression is
\begin{equation}\label{main}
\langle P(f)\rangle_{\rm ratio}=R_{0}\exp\left({\frac{2L}{v_{\rm ph}\xi_{\rm E}}f}\right),
\end{equation}
with $R_{0}=P_{\rm out}(f)/P_{\rm in}(f)$ the ratio of powers generated at the two foot-points. In this equation, $\langle P(f)\rangle_{\rm ratio}$ depends on two factors. The exponential factor depends on wave propagation and damping properties: the wave travel time along the full path, $2L/v_{\rm ph}$, the frequency and the damping ratio. In the presence of damping, because in the shaded region waves in the inward direction are damped to a greater extent than waves in the outward direction this factor is greater than one. Its largest possible value, in the limit of strong damping ($\xi_{\rm E} =L_{\rm D}/\lambda =1/\pi$) is $\exp\left(2L/L_{\rm D}\right)$. In the absence of damping, $\xi_{\rm E} \rightarrow\infty$ and  $\langle P(f)\rangle_{\rm ratio}=R_{\rm 0}$. The exponential factor reduces to one and the average power ratio is equal to the ratio of powers at the two foot-points. 
	
The parameter $R_0$ is our second relevant parameter. If  the power generated at the two foot-points is the same, $R_0=1$, and resonant damping is the only possible contributor to the observed average power ratio. If $R_0\neq 1$, the asymmetry in foot-point driving will be a factor that will either increase  ($P_{\rm out}(f)> P_{\rm in}(f)$, $R_0>1$) or decrease ($P_{\rm out}(f)< P_{\rm in}(f)$, $R_0<1$) the contribution of resonant damping to the average power ratio. In general, both processes will contribute. 

In the following, we consider that $R_0$ is a numerical factor, independent of frequency. We also use $M_{\rm R}$ to denote the reduced model that considers only resonant damping. We use $M_{\rm A}$ to denote the larger model that, in addition, includes the contribution from foot-point wave power asymmetry.

	\section{Analysis and results}
	In this section, we use average power ratio measurements to infer information about the relevant physical parameters, $\xi_{\rm E}$ and $R_0$. This information is then used to assess the evidence for the two considered models. The forward problem is given by Eq.~(\ref{main}). Our data are the set of CoMP measurements analysed by \cite{Verth2010}. They are displayed as triangles in Fig.~\ref{f2}. We use Bayesian methods for solving both inference and model comparison problems \citep[see e.g.,][]{Arregui2018}.
	
	\subsection{Inference}\label{inference}
	
	In Bayesian inference the marginal posterior for a given parameter $\theta_{k}$ conditional on a model $M$ with parameter set 
	$\mbox{\boldmath$\theta$}$ and observed data $d$ is
	\begin{equation}\label{eq6}
	p(\theta_k|M,d)=\int{p(\mbox{\boldmath$\theta$}|M,d)d\theta_1\,...\,d\theta_{k-1}d\theta_{k+1}\,...\,d\theta_n}. 
	\end{equation}
	In this expression, $p(\mbox{\boldmath$\theta$}|M,d)$ is the global posterior obtained from the Bayes' theorem.
	
	We first perform the inference by considering individually the measured average power ratio values at each frequency, hence $d= d_{\rm i}= \langle P(f_{\rm i})\rangle_{\rm ratio}$. A Gaussian likelihood is used to compute the global posterior. We adopt an error model with a 10\% relative error on average power ratio to stand for the increase in the noise of CoMP data with frequency \citep{Tomczyk2009,Verth2010}. This is commensurate with the estimates by \citet{Tomczyk2009} and \citet{Morton2015}. The reduced model ($M_{\rm R}$) contains three parameters, $\mbox{\boldmath$\theta$}_{\rm R}=\{ L,v_{\rm ph},\xi_{\rm E}\}$. The large model ($M_{\rm A}$) has four parameters, $\mbox{\boldmath$\theta$}_{\rm A}=\{ R_0,L,v_{\rm ph},\xi_{\rm E}\}$ and includes the reduced model as a particular case (when $R_0=1$). From the estimates of \cite{Tomczyk2009}, we adopt Gaussian priors for $L$ and $v_{\rm ph}$ centred at 250 Mm and 600 km s$^{-1}$, respectively. We assume 10\% standard deviations from the mean in both cases, comparable to the values given by \citet{Morton2015} and \citet{Tiwari2019}. For the damping ratio and the power ratio at the foot-points, we initially chose broad enough uniform priors to accommodate
	our uncertain posteriors, $\mathcal{U} (\xi_{\rm E},1/\pi,10)$ and $\mathcal{U} (R_0,0.1,4)$, respectively.
	
\begin{figure}[!t]
		\includegraphics[scale=0.572]{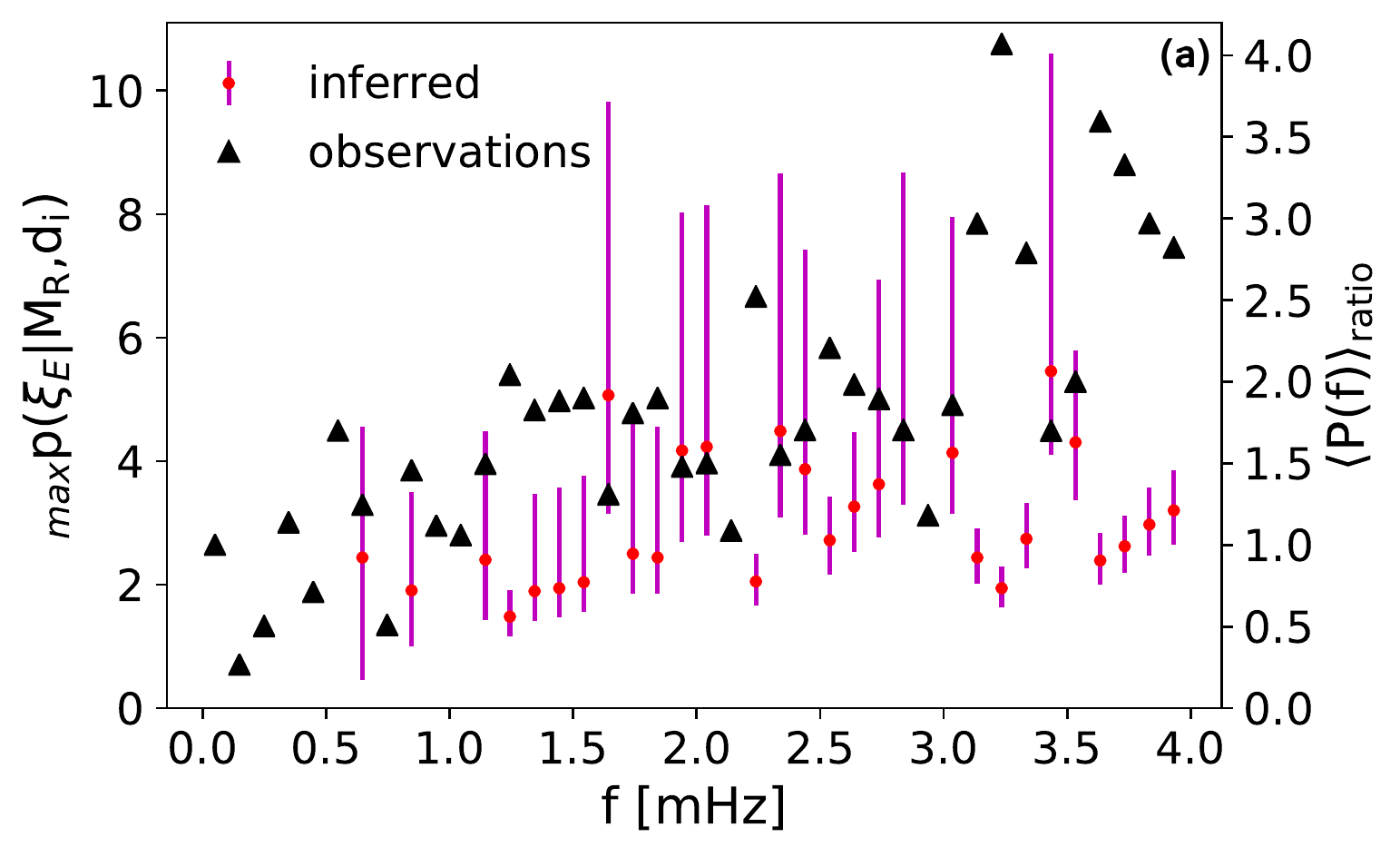} 	
		\includegraphics[scale=0.572]{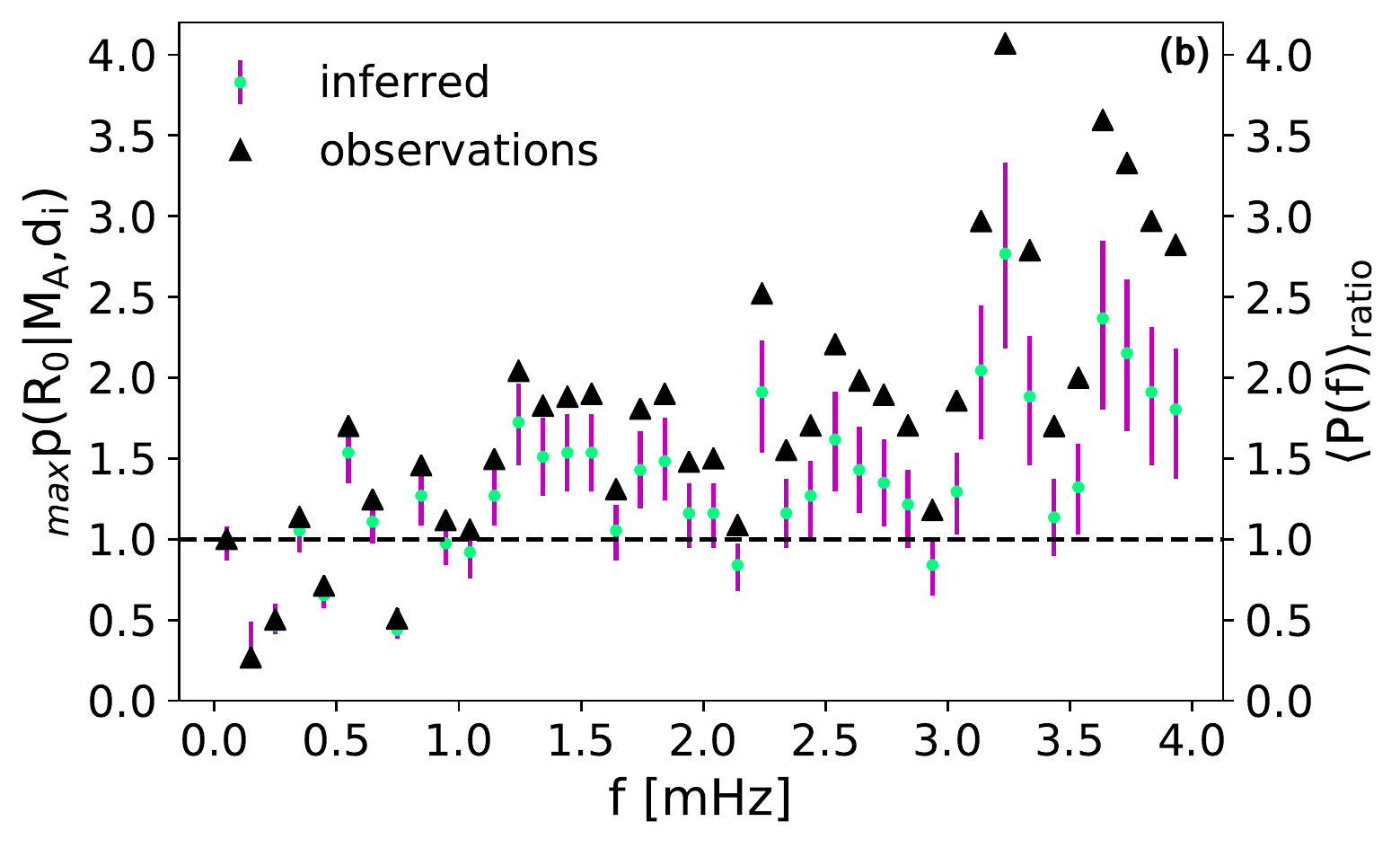}	
		\includegraphics[scale=0.572]{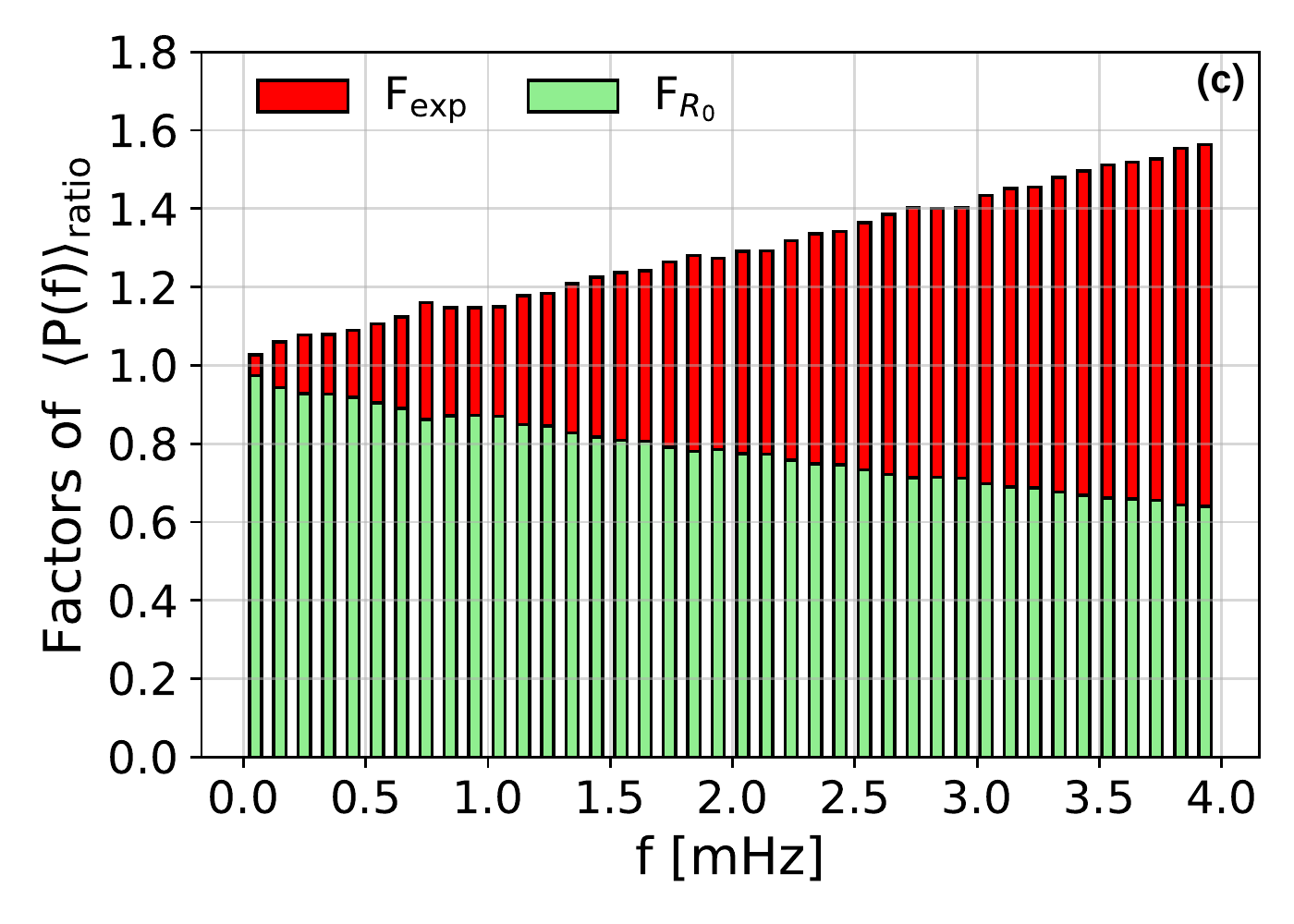}
		\caption{ (a) and (b) Inferences of damping ratio and power ratio at foot-points using CoMP data for average power ratio at different frequencies (triangles). The red and green dots are the maximum a posteriori estimates with uncertainty at the 68\% credible interval (purple bars). In (a), the inference is restricted to $\langle P(f)\rangle_{\rm ratio}>1.2$. In (b), the horizontal dashed line is at $\langle P(f)\rangle_{\rm ratio}=1$. (c) Estimates of fractional contribution from power ratio at foot-points, ${\rm F}_{\rm R_0}$, and absolute contribution from resonant damping, ${\rm F}_{\rm exp}$. \label{f2}}
	\end{figure}

	Consider first that resonant damping is the only cause for the observed data. The inference of the parameter vector  
	$\mbox{\boldmath$\theta$}_{\rm R}=\{ L,v_{\rm ph},\xi_{\rm E}\}$ conditional on model $M_{\rm R}$ and data $d_{\rm i}$ is a well determined problem. The use of Gaussian priors for $L$ and $v_{\rm ph}$ makes this inference a one-parameter one-observable problem. Constrained posteriors are obtained for the damping ratio, $\xi_{\rm E}$. Posterior summaries are shown in Fig.~\ref{f2}a.  The inferences shows variability, but no apparent tendency with frequency. The maximum a posteriori estimates are in the range $\hat{\xi}^{\rm MAP}_{\rm E}={\rm max} p(\xi_{\rm E}|M_{\rm R},d_{\rm i}) \sim (1,5)$. Information on data narrows down our original prior assumption. 
	
	Consider next foot-point wave power asymmetry as an additional possible factor. The inference of the parameter vector  
	$\mbox{\boldmath$\theta$}_{\rm A}=\{ R_0,L,v_{\rm ph},\xi_{\rm E}\}$ conditional on model $M_{\rm A}$ and data $d_{\rm i}$ is an underdetermined problem.  In essence, this is a two-parameter one-observable inference. Out of the two relevant parameters, the data are able to constrain the power ratio at the foot-points $R_0$ only.  Posterior summaries are shown in Fig.~\ref{f2}b. The maximum a posteriori estimates are in the range $\hat{R}^{\rm MAP}_{0}={\rm max} p(R_0|M_{\rm A},d_{\rm i})\sim(0.25,2.8)$.
	They are underneath the average power ratio values and their magnitudes constitute a significant fraction of the measured average power ratio. Although estimates for $\hat{\xi}^{\rm MAP}_{\rm E}$ cannot be obtained, Eq.~(\ref{main}) indicates that a balance must exist such that estimates with $\hat{R}^{\rm MAP}_{0} < 1$ are compensated with lower estimates for $\hat{\xi}^{\rm MAP}_{\rm E}$ (stronger damping) and viceversa for $\hat{R}^{\rm MAP}_{0} > 1$.
	
	\begin{figure*}[!t]
		\centering
		\includegraphics[scale=0.6]{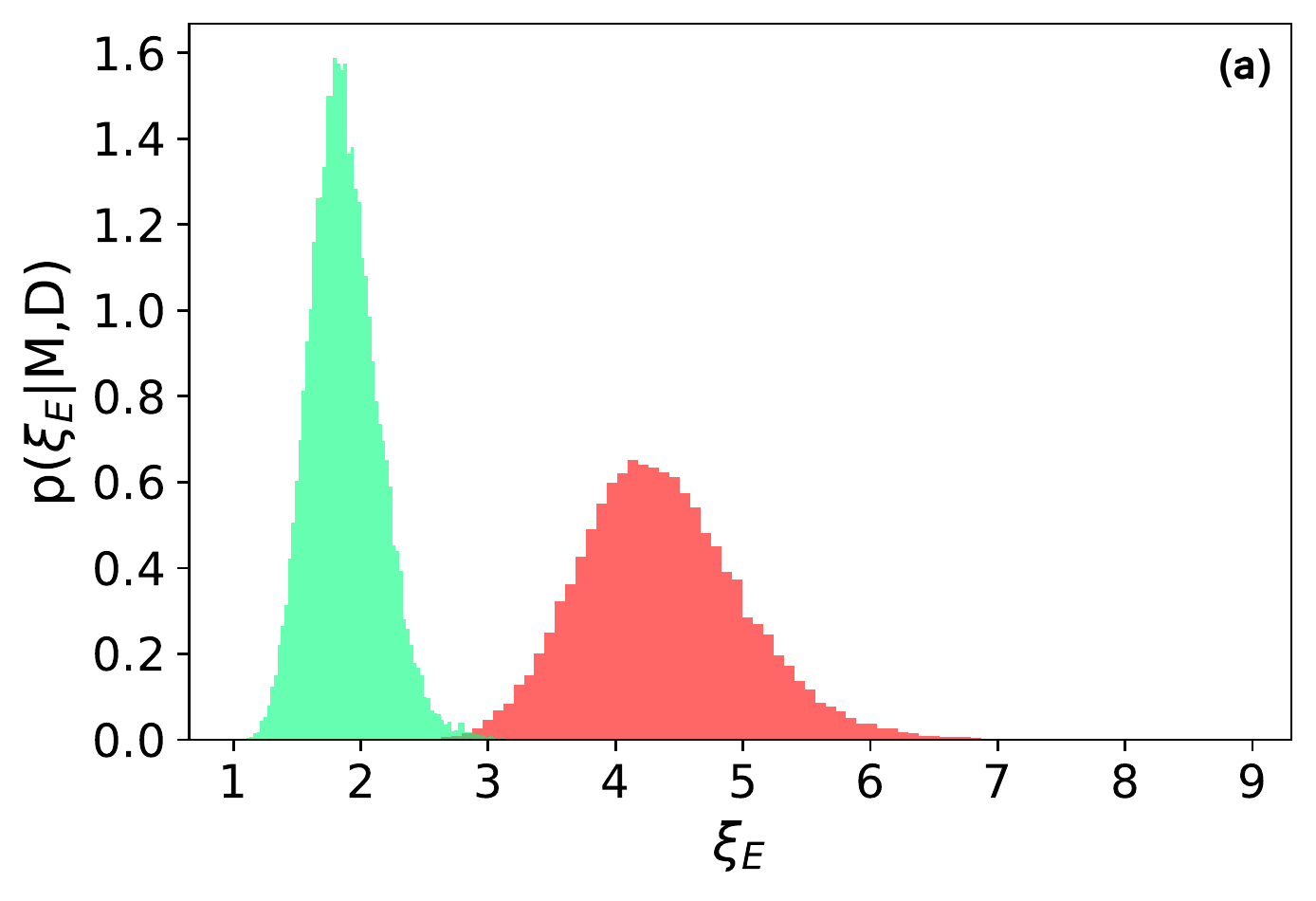}
		\includegraphics[scale=0.6]{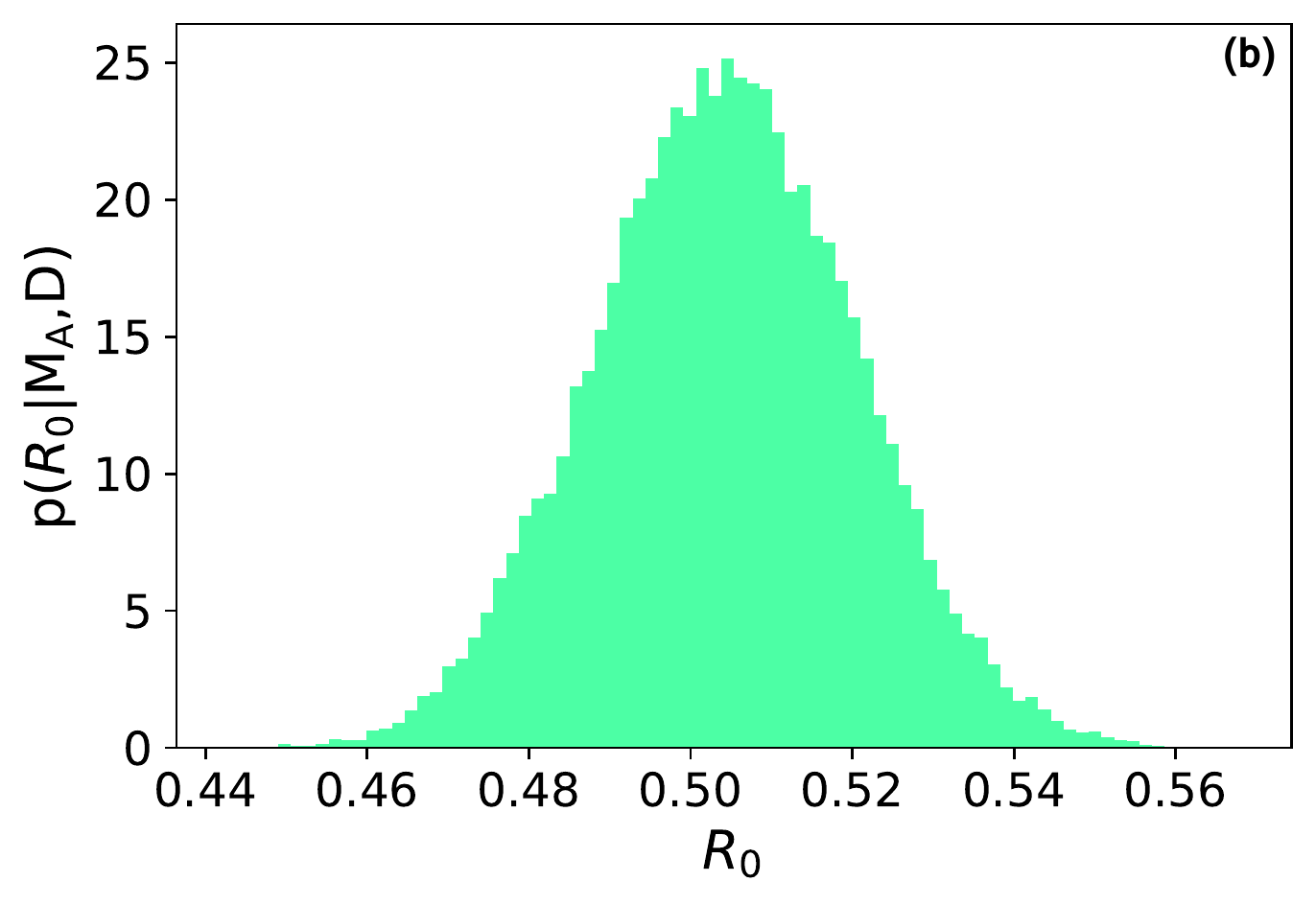}
		\caption{ (a) Marginal posterior distributions for $\xi_{\rm E}$ conditional on data $D$ and models $M_{\rm R}$ (red) and  $M_{\rm A}$ (green). (b) Marginal posterior distribution for $R_{0}$ conditional on data $D$ and model $M_{\rm A}$. The data $D$ consist of the collective use of the CoMP data set analysed by \citet{Verth2010}. The posterior summaries are $\hat{\xi}^{\rm MAP}_{\rm E}=4.3^{+0.7}_{-0.6}$ for $p(\xi_{\rm E}|M_{\rm R},D)$,  $\hat{\xi}^{\rm MAP}_{\rm E}=1.9^{+0.3}_{-0.2}$ for $p(\xi_{\rm E}|M_{\rm A},d)$, and $\hat{R}^{\rm MAP}_{0}=0.5\pm 0.02$ for $p(R_0|M_{\rm A},D)$, with uncertainty given at the 68\% credible interval. The posteriors are computed by Markov Chain Monte Carlo (MCMC) sampling using the Python {\sl emcee} package \citep{emcee}. \label{f3} }
	\end{figure*} 
	
	We can use the $\hat{R}^{\rm MAP}_{0}$ estimates to compute the contributions from the two factors 
	in Eq.~(\ref{main}) to the measured average power ratio.  We define ${\rm F}_{\rm R_0}= \hat{R}^{\rm MAP}_{0} / \langle P(f_{\rm i})\rangle_{\rm ratio}$. It follows that ${\rm F}_{\rm exp} = {\rm F}^{-1}_{R_0}$. Figure~\ref{f2}c shows that the contribution from resonant damping alone, ${\rm F}_{\rm exp}$,  is always greater than one. It increases with frequency, but reaches a modest fraction of the observed average power ratio for all inferences. Conversely, the contribution from foot-point power ratio decreases with frequency, but is always a significant fraction of the observed average power ratio.  
	A narrower prior for $R_0\in[0.1-2]$ leads to the same result, with the exception of the inferences with $R^{\rm max}_0$ close to 2 because the upper limit of the prior is too restrictive to contain the full marginal posterior in those cases. 
	
	We can add more information to the inference process using all at once the full set of CoMP data. Consider then $d=D=\{f_{\rm i},\langle P(f_{\rm i})\rangle_{\rm ratio}\}$ in Eq.~(\ref{eq6}).  The definitions for models, parameters and priors are the same as before.  How the data are employed, individually or collectively,  only affects the likelihood function.  We construct a global likelihood as the product of the individual likelihoods at each frequency. The product of this global likelihood and the priors gives the posterior distributions. 
	
	In Fig.~\ref{f3} we show the resulting marginal posterior distributions for the two parameters of interest, $\xi_{\rm E}$ and $R_0$. 
	The inclusion of additional information enables to properly constrain all marginal posteriors. When analysed collectively, the set of CoMP observations is equally well explained by the reduced model, $M_{\rm R}$, with parameter distribution $p(\xi_{\rm E}|M_{\rm R},D)$  and by the larger model, $M_{\rm A}$, with parameter distributions $p(\xi_{\rm E}|M_{\rm A},D)$ and $p(R_0|M_{\rm A},D)$. The full posterior for $R_0$ is below one, hence $P_{\rm out} < P_{\rm in}$ and there is asymmetry in the power generated at both foot-points. The consequent inference for $\xi_{\rm E}$ is shifted towards stronger damping regimes to compensate the decreasing factor due to the discrepancy at the foot-points.
	
	\subsection{Model comparison}\label{comparison}
	
	Since the two models above are equally good in explaining observed data, we appeal to model comparison.  We quantify their relative plausibility using the Bayes factor
	
	\begin{equation}
	B_{\rm RA}=\frac{p(\mathcal{D}|M_{\rm R})}{p(\mathcal{D}|M_{\rm A})} ,
	\end{equation}
	which, in the case of equal prior belief in both models, is identical to the posterior ratio.  In this expression,
	\begin{equation}\label{marginal}
	p(\mathcal{D}|M_{\rm R,A})=\int_{\mbox{\boldmath$\theta$}_{\rm R,A}}{p(\mbox{\boldmath$\theta$}_{\rm R,A}|M_{\rm R,A})p(\mathcal{D}| \mbox{\boldmath$\theta$}_{\rm R,A}, M_{\rm R,A}) d\mbox{\boldmath$\theta$}_{\rm R,A}},
	\end{equation}
	are the marginal likelihoods for models $M_{\rm R,A}$, a measure of their plausibility that informs on how well the data are predicted by them.
	
	In the above equations, $\mathcal{D}$ are synthetic data in the two-dimensional data space, covering the full ranges in frequency and average power ratio of CoMP observations, $\mathcal{D}=(f, \langle P(f)\rangle_{\rm ratio})$. They are generated using Eq.~(\ref{main}) over a grid of points in $f$ and $\langle P(f)\rangle_{\rm ratio}$. As for the priors, we use those of the inference analysis for $L$ and $v_{\rm ph}$. Our state of knowledge on $L$ and $v_{\rm ph}$ affects in the same manner models $M_{\rm R}$ and $M_{\rm A}$. Because the distributions for the two relevant parameters in Fig.~\ref{f3} are approximately Gaussian, we adopt Gaussian priors with means and standard deviations equal to the inference summaries in Fig.~\ref{f3}.
	
	\begin{figure*}[!h]
		\includegraphics[scale=0.6]{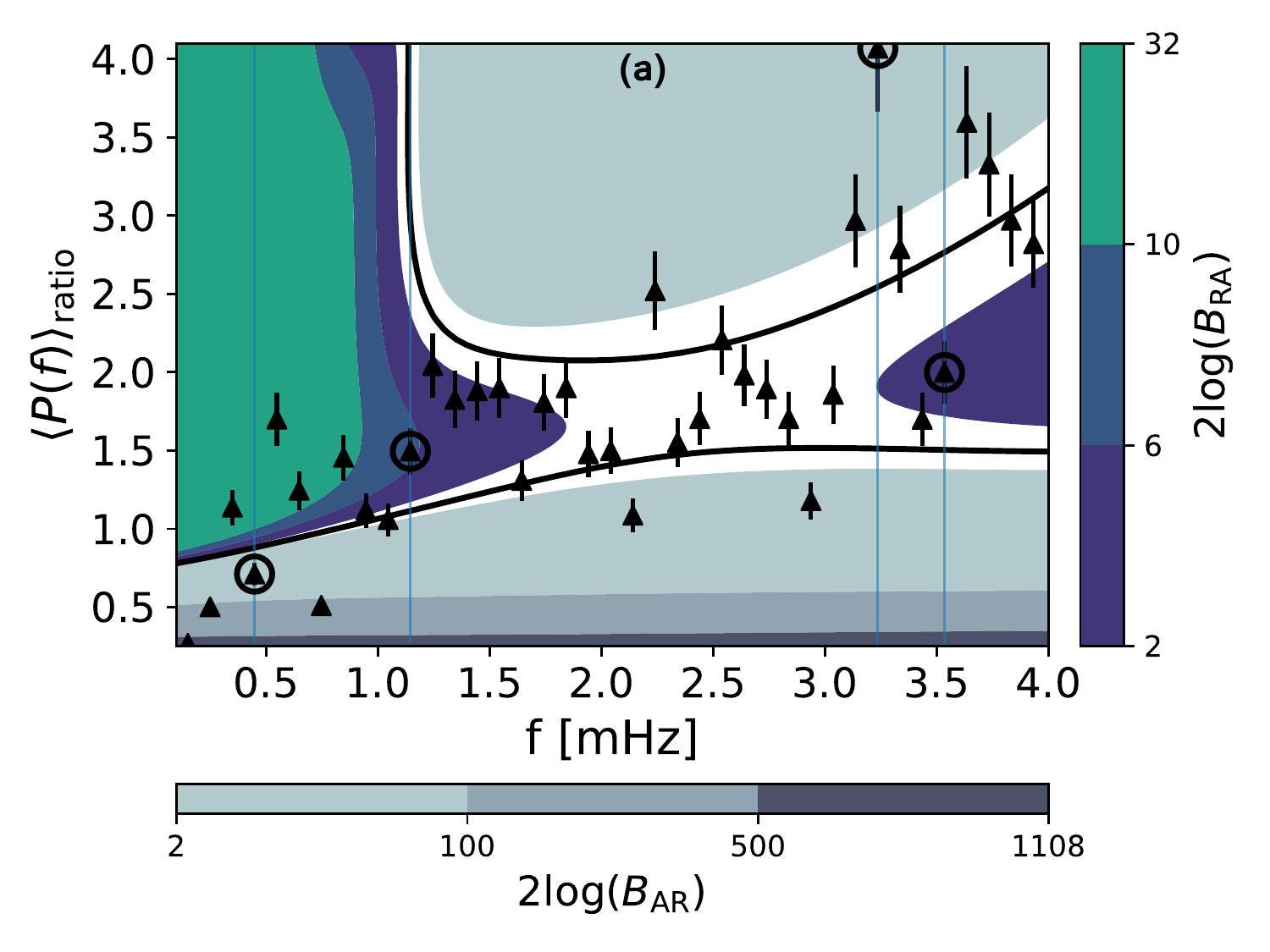}
		\includegraphics[scale=0.6]{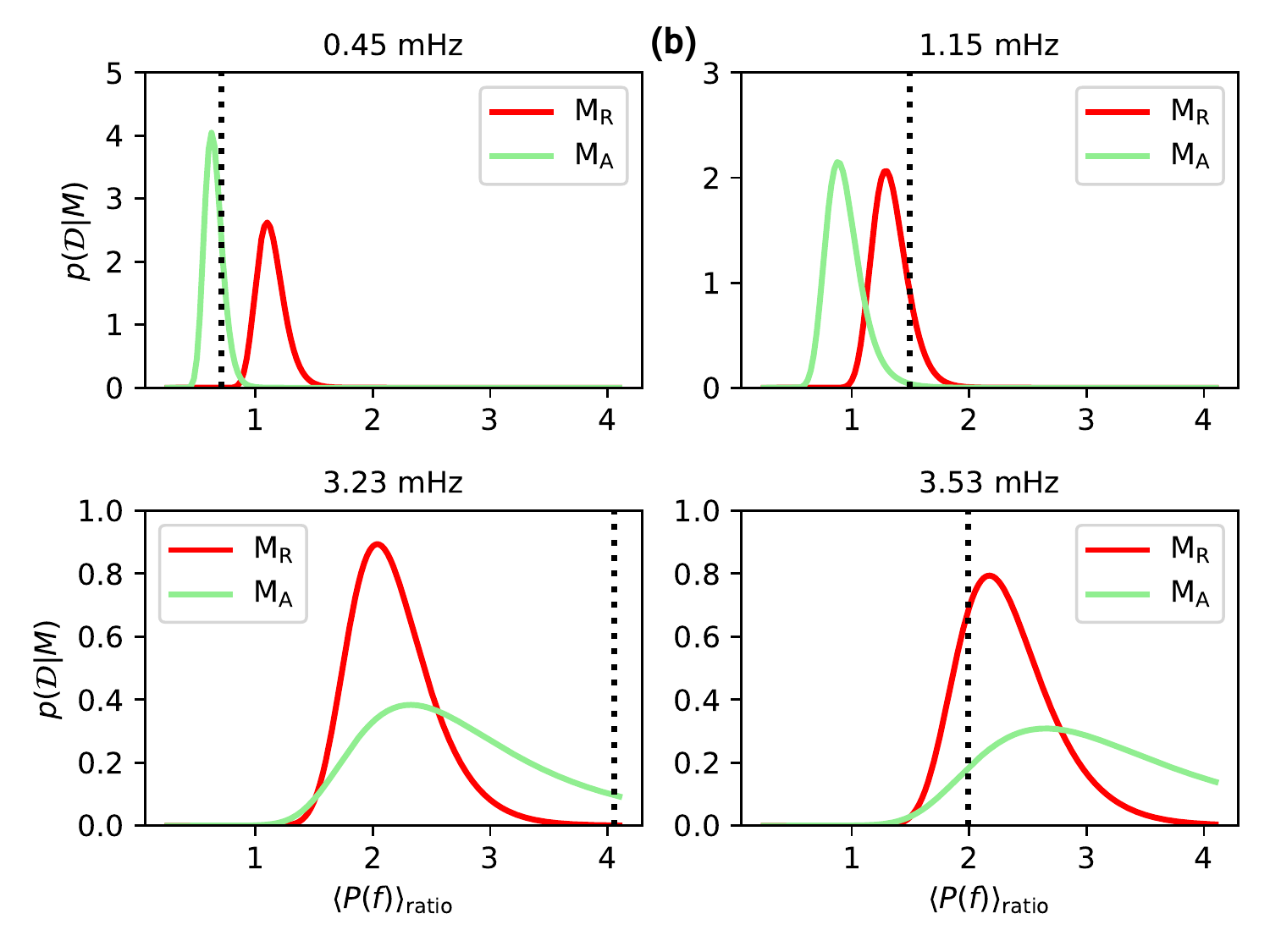}
		\caption{ (a) Filled contour with the distribution of Bayes factors, $B_{\rm RA}$ and $B_{\rm AR}$, over the two-dimensional data space $\mathcal{D}$.  Solid lines connect points with $p(\mathcal{D}|M_{\rm R}) = p(\mathcal{D}|M_{\rm A})$ (Bayes factor zero).   The computations are performed over a grid of points ($N_{f}=80$, $N_{\rm \langle P(f)\rangle_{\rm ratio}}=155$) over the ranges $f\in[0.05,4]$ and $\langle P(f)\rangle_{\rm ratio}\in[0.25,4.1]$. The priors are $p(\xi_{\rm E}) \sim \mathcal{G}(1.9,0.3)$ for $M_{\rm R}$ and  $p(\xi_{\rm E}) \sim \mathcal{G}(4.3,0.6)$ ; $p(R_0) \sim \mathcal{G}(0.5,0.02)$ for $M_{\rm A}$. Triangles represent CoMP data. The error model for observed and synthetic average power ratio is the same adopted in the inference analysis, Sect.~\ref{inference}. Following \citet{Kass1995}, the evidence in favour of a model $i$  in front of an alternative $j$ is inconclusive for values of 2$\log(B_{ij})$ from 0 to 2; positive from 2 to 6; strong from 6 to 10; and very strong for values above 10. (b) Marginal likelihood for models $M_{\rm R}$ and $M_{\rm A}$ for the data points circled in (a) as a function of average power ratio. The vertical dotted lines are the CoMP average power ratio values. \label{f4}}
	\end{figure*}
	
	In Fig.~\ref{f4}a we show the resulting Bayes factors over the two-dimensional data space.
	The distribution of evidence shows three well defined regions delimited by the boundaries where the Bayes factor is zero (equal marginal likelihoods). In the central region, the marginal likelihood is larger for model $M_{\rm R}$ than for model $M_{\rm A}$. Hence, in this region model $M_{\rm R}$ is more plausible than model $M_{\rm A}$. However, levels of evidence depend on the magnitude of the Bayes factor, translated using the classification criteria by \citet{Kass1995}. Based on the magnitude of the Bayes factor $B_{\rm RA}$, the evidence in favour of $M_{\rm R}$ is inconclusive in the white area within this region and varies from positive to very strong in the blue to green areas. We note that in the green area, for frequencies below $\sim$1 mHz and average power ratios above $\sim 2$, the large values of $B_{\rm RA}$ are due to ratios of small marginal likelihood values and are thus not a genuine signature of evidence. None of the observed data falls in this area. In the regions above and below the solid lines, the marginal likelihood is larger for model $M_{\rm A}$ than for model $M_{\rm R}$. Hence, in these two regions model $M_{\rm A}$ is more plausible than model $M_{\rm R}$. Based on the magnitude of the Bayes factor $B_{\rm AR}$, the evidence in favour of $M_{\rm A}$ is inconclusive in the white areas within these two regions and varies from positive to very strong as we move to the upper and lower areas, respectively.
		
	Observed CoMP data with the assumed error bars are superimposed over the evidence distribution. In most of the cases, data fall over regions where the marginal likelihood for model $M_{\rm R}$ is larger. However, only a fraction of them falls over areas where the evidence supports model $M_{\rm R}$ in front of model $M_{\rm A}$. A considerable proportion of them falls into the two regions where the marginal likelihood for model $M_{\rm A}$ is larger. In a non-negligible number of cases, they are over areas where the evidence supports model $M_{\rm A}$ in front of model $M_{\rm R}$.
	
	We plot in Fig.~\ref{f4}b the marginal likelihood for models $M_{\rm R,A}$ for four representative data points as a function of synthetic average power ratio. The relative magnitude of the two marginal likelihoods at the measured average power ratio determines the Bayes factors, hence the evidence.
	
	Consider data at $0.45$ mHz with average power ratio 0.71. At that frequency, model $M_{\rm R}$ predicts average power ratios that are above the measured value making model $M_{\rm A}$ more plausible. On the other hand, measured average power ratios at $1.15$ and $3.53$ mHz are better explained by model $M_{\rm R}$ with strong and positive evidence, respectively.  Data at 3.23 mHz represent a singular example. Because the measured average power ratio is much larger than those well predicted by model $M_{\rm R}$, the evidence favours model $M_{\rm A}$. Similar analyses explain the full distribution of evidence in Fig.~\ref{f4}a.
	
	Moving from low to high frequencies, the marginal likelihoods for both models shift towards larger average power ratio values.  They become distributed over a larger data space. Consequently, their magnitude decreases, following the principle that any prior space that is not supported by data will reduce the evidence (Occam's razor).

\section{Conclusions}\label{conclusions}
 
	In addition to resonant damping, an asymmetry in the wave power generated at the foot-points of propagating coronal waves can contribute to CoMP measurements of integrated average power ratio. The increasing  or decreasing character of the contribution is determined by the relative amplitude of the waves generated at the two foot-points connecting the wave path. 
	
	The observations are equally well explained by reduced models with identical power at the two foot-points and by larger models with foot-point wave power asymmetry ($R_0\neq1$) and relatively weaker or stronger damping.  If $R_0 < 1$, the inference with the reduced model overestimates the damping ratio. If $R_0 >1$, the inference with the reduced model underestimates the damping ratio. The decision on which model to use affects the inferences about the strength of the damping and, thus, has implications on the assessment of wave-based coronal heating.
	
	For the particular dataset first analysed by \cite{Verth2010} and the errors adopted, inference results at different frequencies show that the contribution is more important for low frequency waves, which are less affected by resonant damping.  A global inference analysis results in estimates equivalent to a value of $\sim 0.7$ for the wave amplitude ratio between the two foot-points connecting the wave path. 
	
	Adding complexity to the reduced model is justified. Both models explain the data with different levels of plausibility, depending on the particular combination of measured frequency and average power ratio. The larger model is more complex and can, thus, make more predictions. But they should be supported by data. Our distribution of evidence indicates that this is the case. A sizeable proportion of data, with average power ratio values below and above those well predicted by the reduced model, are better explained by the larger model.
	
	Our analysis was limited to a particular dataset along a single wave path, the consideration of $R_0$ as a numerical factor, and the assumption of a particular error model for average power ratio data. Further advancement can result from a systematic application of our methods to existing CoMP datasets \citep[e.g.][]{Morton2015,Tiwari2019}, further observational characterisation of $R_0$, with its possible frequency dependence, and the increase in precision in Doppler imaging of the corona.

\begin{acknowledgements}
	We acknowledge financial support from the Spanish Ministerio de Ciencia, Innovaci\'on y Universidades through project PGC2018-102108-B-I00 and from FEDER funds. M.M-S. acknowledges financial support through a Severo Ochoa FPI Fellowship under the project SEV-2015-0548.  The data employed in this study were obtained by \citet{Tomczyk2007} using the Coronal Multi-Chanel Polarimeter (CoMP) instrument and kindly provided by G. Verth.
\end{acknowledgements}

\bibliographystyle{aa} 
\bibliography{biblio} 

\begin{thebibliography}{17}
\expandafter\ifx\csname natexlab\endcsname\relax\def\natexlab#1{#1}\fi

\bibitem[{Arregui(2015)}]{Arregui2015}
Arregui, I. 2015, Royal Society of London Philosophical Transactions Series A,
  373, 20140261; DOI: 10.1098/rsta.2014.0261

\bibitem[{{Arregui}(2018)}]{Arregui2018}
{Arregui}, I. 2018, Advances in Space Research, 61, 655

\bibitem[{{Foreman-Mackey} {et~al.}(2013){Foreman-Mackey}, {Hogg}, {Lang}, \&
  {Goodman}}]{emcee}
{Foreman-Mackey}, D., {Hogg}, D.~W., {Lang}, D., \& {Goodman}, J. 2013, \pasp,
  125, 306

\bibitem[{{Goossens} {et~al.}(2012){Goossens}, {Andries}, {Soler}, {Van
  Doorsselaere}, {Arregui}, \& {Terradas}}]{Goossens2012a}
{Goossens}, M., {Andries}, J., {Soler}, R., {et~al.} 2012, \apj, 753, 111

\bibitem[{{Jess} {et~al.}(2009){Jess}, {Mathioudakis}, {Erd{\'e}lyi},
  {Crockett}, {Keenan}, \& {Christian}}]{Jess2009}
{Jess}, D.~B., {Mathioudakis}, M., {Erd{\'e}lyi}, R., {et~al.} 2009, Science,
  323, 1582

\bibitem[{{Kass} \& {Raftery}(1995)}]{Kass1995}
{Kass}, R.~E. \& {Raftery}, A.~E. 1995, JASA, 90, 773

\bibitem[{{Liu} {et~al.}(2019){Liu}, {Nelson}, {Snow}, {Wang}, \&
  {Erd{\'e}lyi}}]{Liu2019}
{Liu}, J., {Nelson}, C.~J., {Snow}, B., {Wang}, Y., \& {Erd{\'e}lyi}, R. 2019,
  Nature Communications, 10, 3504

\bibitem[{{McIntosh} \& {De Pontieu}(2012)}]{McIntosh2012}
{McIntosh}, S.~W. \& {De Pontieu}, B. 2012, \apj, 761, 138

\bibitem[{Montes-Sol{\'\i}s \& Arregui(2017)}]{Yo2017}
Montes-Sol{\'\i}s, M. \& Arregui, I. 2017, The Astrophysical Journal, 846, 89

\bibitem[{{Morton} {et~al.}(2015){Morton}, {Tomczyk}, \& {Pinto}}]{Morton2015}
{Morton}, R.~J., {Tomczyk}, S., \& {Pinto}, R. 2015, Nature Communications, 6,
  7813

\bibitem[{{Pant} {et~al.}(2019){Pant}, {Magyar}, {Van Doorsselaere}, \&
  {Morton}}]{Pant2019}
{Pant}, V., {Magyar}, N., {Van Doorsselaere}, T., \& {Morton}, R.~J. 2019,
  \apj, 881, 95

\bibitem[{{Srivastava} {et~al.}(2017){Srivastava}, {Shetye}, {Murawski},
  {Doyle}, {Stangalini}, {Scullion}, {Ray}, {W{\'o}jcik}, \&
  {Dwivedi}}]{Srivastava2017}
{Srivastava}, A.~K., {Shetye}, J., {Murawski}, K., {et~al.} 2017, Scientific
  Reports, 7, 43147

\bibitem[{{Terradas} {et~al.}(2010){Terradas}, {Goossens}, \&
  {Verth}}]{Terradas2010}
{Terradas}, J., {Goossens}, M., \& {Verth}, G. 2010, \aap, 524, A23

\bibitem[{{Tiwari} {et~al.}(2019){Tiwari}, {Morton}, {R{\'e}gnier}, \&
  {McLaughlin}}]{Tiwari2019}
{Tiwari}, A.~K., {Morton}, R.~J., {R{\'e}gnier}, S., \& {McLaughlin}, J.~A.
  2019, \apj, 876, 106

\bibitem[{{Tomczyk} \& {McIntosh}(2009)}]{Tomczyk2009}
{Tomczyk}, S. \& {McIntosh}, S.~W. 2009, \apj, 697, 1384

\bibitem[{{Tomczyk} {et~al.}(2007){Tomczyk}, {McIntosh}, {Keil}, {Judge},
  {Schad}, {Seeley}, \& {Edmondson}}]{Tomczyk2007}
{Tomczyk}, S., {McIntosh}, S.~W., {Keil}, S.~L., {et~al.} 2007, Science, 317,
  1192

\bibitem[{{Verth} {et~al.}(2010){Verth}, {Terradas}, \& {Goossens}}]{Verth2010}
{Verth}, G., {Terradas}, J., \& {Goossens}, M. 2010, \apjl, 718, L102

\end{thebibliography}

\end{document}